\documentclass[twocolumn,prl,aps]{revtex4}
\usepackage{graphicx}

\newcommand{\be}{\begin{equation}}
\newcommand{\ee}{\end{equation}}
\newcommand{\bea}{\begin{eqnarray}}
\newcommand{\eea}{\end{eqnarray}}
\newcommand{\HH}{{\cal H}}

\newcommand{\la}{\langle}
\newcommand{\ra}{\rangle}

\newcommand{\lp}{\left(}
\newcommand{\rp}{\right)}
\renewcommand{\vec}[1]{{\bf #1}}

\begin{document}
\title{Excitation of the Dissipationless Higgs Mode in
a Fermionic Condensate}
%% at Strong Coupling}
\author{R. A. Barankov$^1$ and L. S. Levitov$^2$}
\affiliation{ $^1$ Department of Physics, University of Illinois, 1110 West Green Street, Urbana, IL 61801
\\$^2$Department of Physics,
Massachusetts Institute of Technology, 77 Massachusetts Ave, Cambridge, MA
02139}

\begin{abstract}
The amplitude mode of a fermionic superfluid, analogous to the Higgs Boson,
becomes undamped in the strong coupling regime when its frequency is pushed
inside the BCS energy gap. We argue that this is the case
%happens
in cold gases due to the energy dispersion and nonlocality
of the pairing interaction,
and propose to use the Feshbach resonance regime for parametric excitation of this mode.
The results presented for the BCS pairing dynamics
%with dispersive interaction
indicate that even weak dispersion suppresses dephasing
and gives rise to persistent oscillations.
%% of this mode.
The frequency of oscillations extracted from our simulation
of the BCS dynamics agrees with the prediction of the many-body theory.
\end{abstract} \pacs{} \keywords{}

\maketitle

The observation of resonance superfluidity in cold atomic
Fermi gases\,\cite{Regal04,Zwierlein04} at magnetically
tunable Feshbach resonances\,\cite{Timmermans99}
opened new avenue of exploring the many-body
phenomena. Similar to the earlier work on cold Bose gases
which triggered studies of fascinating collective
phenomena\,\cite{Dalfovo99,Pethick_Smith}, fermionic pairing
at Feshbach resonances\,\cite{Holland01,Ohashi03,Bruun04,Ho04,Schwenk05,Chen05}
presents new opportunities.
In particular, the high degree of coherence
of trapped atoms,
and the possibility to control particle interaction {\it in situ} on
the shortest collective time-scale,
the inverse Fermi energy~\cite{Zwierlein05}, can facilitate exploring
new regimes which are difficult to realize in solid state systems.

The theory of fermionic pairing predicts two principal collective modes
intrinsic to the condensed state.
One is the massless Bogoliubov-Anderson mode
related to the order parameter phase dynamics. Being a Goldstone mode,
it manifests itself in hydrodynamics in the same way as in Bose
systems, and was probed recently in the experiments on
gas expansion and oscillation in traps\,\cite{Bourdel03,Ferlaino04,Kinast05,Stewart06}.
In addition, there exists a second fundamental elementary
excitation\,\cite{Vaks61,Bardasis61,Varma02},
related to the dynamics of the order parameter modulus $|\Delta|$.
Notably, this excitation is unique to fermionic pairing
and has no counterpart in Bose systems\,\cite{Varma02}.
This massive excitation, characterized by a finite frequency, is analogous
to the Higgs Boson in particle physics.
Like the latter it remained elusive, for a long time evading direct probes,
although some indirect manifestations have been discussed\,\cite{Ivlev72,Littlewood82}.
The main obstacle to the detection of the Higgs mode in superconductors is
that it is essentially decoupled from the phase mode responsible
for hydrodynamics and superfluidity.

\begin{figure}[t]
\includegraphics[width=3.5in]{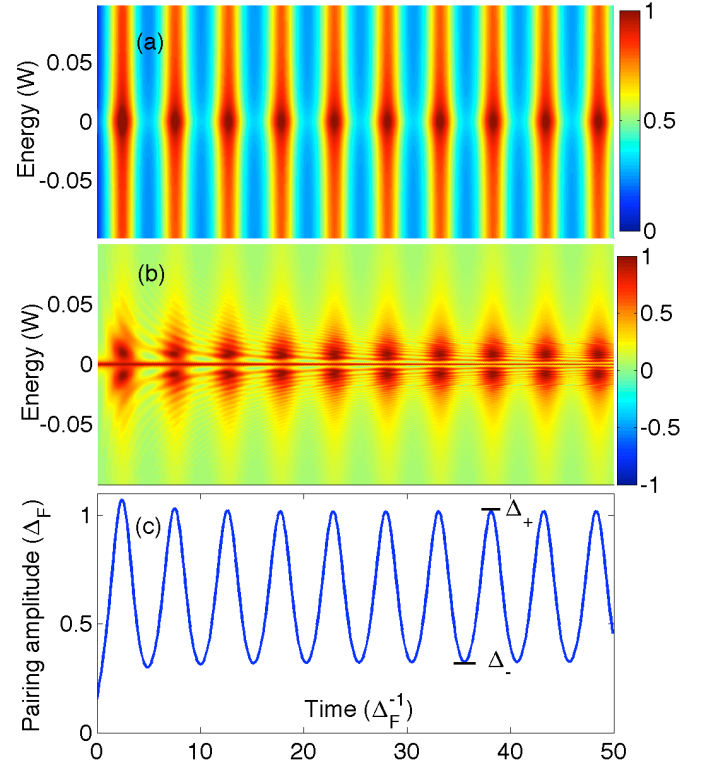}
%\vspace{-0.25cm}
\vspace{-0.8cm}
\caption[]{Non-decaying Higgs mode in a Fermi gas with energy-dependent
pairing interaction
excited by the interaction switching from $g_i$ at $t<0$ to $g$ at $t>0$.
Shown are the time and energy dependence of the pairing amplitude (a),
the $x$-component of the pseudospin vector (b),
and the pairing amplitude at the Fermi energy (c)
as obtained from the model (\ref{eq:interaction}),(\ref{eq:Bloch})
at $g=0.43$, $g_i=0.23$, $a_1=a_2=0.5$, $\gamma/W=0.01$, $\Delta_F/W=0.016$.
Note the initial transient of few periods,
exhibiting some dephasing in pseudospin dynamics (b),
followed by synchronized collective oscillations of fermion states.
%%after which individual fermion states synchronize.
} \vspace{-0.3cm} \label{fig:A_regime}
\end{figure}

In this work we propose to use the dynamical control of pairing interaction demonstrated in
Refs.\cite{Regal04,Zwierlein04} for parametric excitation of the Higgs mode.
We argue that fermion superfluidity in the strong coupling regime realized near
Feshbach resonance represents a distinct advantage, since in this case the Higgs
mode is pushed inside the superconducting gap, $\hbar\omega<2\Delta$,
which eliminates damping due to coupling to quasiparticles.
We demonstrate that this mode can be excited
by a time-dependent pairing interaction, as illustrated in Fig.\,\ref{fig:A_regime}.
In contrast, the BCS theory at weak coupling predicts the Higgs mode
frequency right at the edge of the quasiparticle continuum,
$\hbar\omega= 2\Delta$\,\cite{Varma02},
which leads to collisionless damping of this mode\,\cite{Volkov74,Yuzbashyan05-2}.

%%%%%%%%%%%%%%%%%%%%%%%%%%%%%%%%%%%%%%%%%%%%%%%%%%%%%%%%%%%%%%%%%
%%%%%%%%%%%%%%%%%%%%%%%%%%%%%%%%%%%%%%%%%%%%%%%%%%%%%%%%%%%%%%%%%%

The departure from the behavior at weak coupling arises from the change in the character
of pairing interaction in the strong coupling regime, in particular due to its finite spatial radius
and frequency dispersion. Spatial nonlocality of pairing interaction is known to lead
to discrete collective modes inside the BCS gap\,\cite{Vaks61,Bardasis61}.
Similarly, the energy dispersion of the pairing interaction
and pairing amplitude $\Delta_{\bf p}$ that
becomes prominent at strong coupling\,\cite{Eliashberg60},
leads to discrete collective mode spectrum (see below). While the exact form
of this dispersion
is sensitive to the specifics of the strong coupling problem, it is established in the
literature that, generally, both effects can occur
near Feshbach resonance\,\cite{Szymanska05,Ohashi03,Bruun04,Schwenk05,Chen05}.
Although our understanding of the detailed microscopic picture may be hampered by the nonpertubative nature of the strong coupling problem,
we shall see that within a simplified model used below
the inequality $\hbar\omega<2\Delta$ is fulfilled under very general conditions.

The dissipationless BCS dynamics\,\cite{Volkov74,Barankov04-1}
and
%its
the possibility
%realization
to realize it
in cold gases\,\cite{Andreev04} attracted much attention
recently\,\cite{Szymanska05,Warner05,Yuzbashyan05-2}.
These investigations, with the exception of Ref.\,\cite{Szymanska05},
focused on the case of pairing interaction which is constant
in the entire fermion energy band, concluding\,\cite{Barankov06-2,Yuzbashyan06} that
several interesting dynamical states, synchronized and desynchronized (or dephased),
can be realized by a sudden change in the interaction strength (see the phase diagram
in Ref.\,\cite{Barankov06-2}).

In contrast, as we shall see below, the dephased behavior is suppressed
in the strong coupling regime when due to the energy dispersion
of the pairing interaction the Higgs mode falls inside the BCS energy gap.
Under these conditions an undamped Higgs mode can be
excited upon a sudden change in interaction.
% This behavior is found to be of generic character, true for factorizable
% pairing interaction, as well as for a non-factorizable interaction.
By analyzing the limit when the interaction dispersion disappears
we show how the different regimes of Ref.\,\cite{Barankov06-2} are recovered.
This correspondence suggests an interpretation of the dephased oscillations
discussed in Refs.\,\cite{Volkov74,Yuzbashyan06,Barankov06-2} as a manifestation of the
Higgs mode, algebraically dephased at $\hbar\omega=2\Delta$.

We shall analyze the pairing dynamics in a spatially uniform system
using the pseudospin representation\,\cite{Barankov04-1} of the BCS problem
in which spin $1/2$ operators $s^{\pm}_\vec p=s^x_\vec p\pm is^y_\vec p$
describe Cooper pairs $(\vec p,-\vec p)$:
\be\label{eq:Hspin}
\HH = -\sum_{\vec p} 2\epsilon_{\vec p} s^z_\vec p-\sum_{\vec p\vec
q}\lambda_{\vec p\vec q}(t) s^-_\vec p s^+_\vec q,
\ee
where $\epsilon_\vec p$ is the free particle spectrum.
The interaction $\lambda_{\vec p\vec q}(t)$
that models the energy dispersion at strong coupling is taken in the form
of a sum of a dispersing and nondispersing parts
\be\label{eq:interaction}
\lambda_{\vec p\vec q}(t)=\frac{g(t)}{\nu_F}\lp a_1+a_2 f_\vec p f_\vec q\rp
,\quad f_\vec p=\frac{\gamma}{\sqrt{\gamma^2+\epsilon_\vec p^2}},
\ee
where the dimensionless
parameter $g(t)$ specifies the interaction time-dependence, the constants
$a_{1,2}\ge 0$ satisfy $a_1+a_2=1$, and $\nu_F$ is the density of states at the Fermi level.
The second term in (\ref{eq:interaction}) features dispersion on the energy scale $\gamma$.
Our motivation for choosing the model (\ref{eq:interaction})
was two-fold. Firstly, the form (\ref{eq:interaction}) is general enough to
provide insight into the role of different features, such as
the energy dispersion
(which is controlled by the parameter $\gamma$) and separability
(which is absent unless $a_1$ or $a_2$ vanishes).
Secondly, our numerical method utilized the rank two form
of (\ref{eq:interaction}), allowing for substantial speedup
that could not be implemented for a more general interaction $\lambda_{\vec p\vec q}$.
In addition, the model (\ref{eq:interaction}) is physically motivated by
the theory of BCS pairing
in the simultaneous presence of a retarded and
non-retarded interaction\,\cite{AndersonMorel}.

Within the mean-field approximation, the dynamical equations derived from Eq.(\ref{eq:Hspin}) assume a Bloch form:
\be\label{eq:Bloch}
\frac{d\vec r_\vec p}{dt}=2\vec b_\vec p\times \vec r_\vec p, \quad \vec b_\vec
p=-(\Delta_\vec p^x,\Delta_\vec p^y,\epsilon_\vec p),
\ee
where $\vec r_\vec p=2\la \vec s_\vec p\ra$ are Bloch vectors, and the
effective magnetic field $\vec b_\vec p$
depends on the pairing amplitude $\Delta_\vec p$. The latter is defined
self-consistently:
\be\label{eq:Delta_def}
\Delta_\vec p=\Delta_\vec p^x+i\Delta_\vec p^y=\sum\limits_{\vec q}\frac{\lambda_{\vec p\vec q}(t)}{2}r^+_\vec q,
\,r^+_\vec p=r^x_\vec p+i r^y_\vec p . \ee
The interaction time dependence of interest is a step-like change
from the initial value $g_i$ to the final value $g$. Without loss of generality,
the phase of the order parameter can be chosen equal zero,
allowing us to consider only
the $x$-component of the pairing amplitude, $\Delta_\vec p=\Delta^x_\vec p$.
As an initial state we take the paired ground state
\be\label{eq:BCS_initial}
r_\vec p^x(0)=\frac{\Delta^i_\vec p}{\sqrt{(\Delta^i_\vec p)^2+\epsilon_\vec p^2}},\quad
r_\vec p^z(0)= \frac{\epsilon_\vec p}{\sqrt{(\Delta^i_\vec p)^2+\epsilon_\vec p^2}} .
\ee
The equilibrium energy-dependent amplitude $\Delta_\vec p$ is determined by
the self-consistency equation
\be\label{Eq:Gap_eq}
\Delta_\vec p=\frac{1}{2}\sum_\vec q \lambda_{\vec p\vec q}\frac{\Delta_\vec q}{\sqrt{\epsilon_\vec q^2+\Delta_\vec q^2}},
\ee
in which $\lambda_{\vec p\vec q}$ is given by (\ref{eq:interaction})
with the parameter values $g_i$ and $g$ for the initial and final state.
The corresponding equilibrium pairing gap values, $\Delta^i_\vec p$ and $\Delta_\vec p$,
are found by numerically solving the integral equation (\ref{Eq:Gap_eq}).
Throughout the paper we use the equilibrium value of the pairing gap at the Fermi level, $\Delta_F$,
at the final coupling $g$ as a natural energy scale to parameterize the dynamics.

We integrate Eqs.(\ref{eq:Bloch}) using the Runge-Kutta method of the $4$-th order
with a time step adjusted to achieve sufficient precision of the calculation.
In our simulation we use $N=10^4,10^5$ equally spaced energy states within
bandwidth $W$, $-W/2<\epsilon_\vec p<W/2$, with the level spacing much
smaller than all other energy scales in the problem.

\begin{figure}[t]
\includegraphics[width=3.5in]{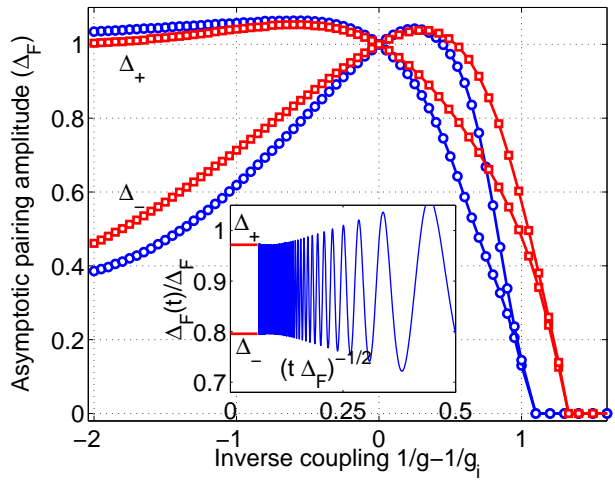}
%\vspace{-0.25cm}
\vspace{-0.8cm} \caption[]{Long-time behavior of $\Delta_F(t)$, the pairing
amplitude at the Fermi level, oscillating between $\Delta_+$ and $\Delta_-$.
Shown are two examples of $\Delta_\pm$ as a function of the initial state for a
non-separable (circles) and a separable (squares) interaction
(\ref{eq:interaction}). Parameters used: $a_1=a_2=0.5$, $g=0.43$,
$\Delta_F/W=0.016$, $\gamma/W=0.01$, and $a_1=0$, $a_2=1$, $g=0.61$,
$\Delta_F/W=0.005$, $\gamma/W=0.01$, respectively. {\it Inset:} Linear fit of a
sample trace $\Delta_F(t)$ {\it vs.} $t^{-1/2}$ used to extract $\Delta_\pm$.}
\vspace{-0.3cm} \label{fig:asymp_diag_narrow}
\end{figure}

We analyze the quantity $\Delta_F(t)$ which at long times oscillates between
the maximum and minimum values $\Delta_+$ and $\Delta_-$.
To find the asymptotic values $\Delta_\pm$ we employ the numerical procedure
sketched in Fig.~\ref{fig:asymp_diag_narrow} inset:
$\Delta_\pm$ are obtained from the linear fits
to the maxima and minima of $\Delta_F$ {\it vs.} $t^{-1/2}$
intersection with the $y$-axis.
The $t^{-1/2}$ time parameterization is motivated by the dephasing law
$\delta\Delta(t)\propto t^{-1/2}$ found in Refs.\cite{Volkov74,Yuzbashyan05-2}
for the energy-independent interaction.
Should the dephasing occur, the asymptotic values would coincide, $\Delta_+=\Delta_-$.

%%%%%%%%%%%%%%%%%%%%%%%%%%%%%%%%%%%%%%%%%%%%%%%%%%%%%%%%%%%%%%%%%
%%%%%%%%%%%%%%%%%%%%%%%%%%%%%%%%%%%%%%%%%%%%%%%%%%%%%%%%%%%%%%%%%%

In contrast to the above, for the interaction (\ref{eq:interaction})
% with Refs.\cite{Volkov74,Yuzbashyan05-2},
the dephased behavior is suppressed. Instead, as illustrated in Fig.~\ref{fig:asymp_diag_narrow}, we observe non-decaying
periodic oscillations
%% collective mode of the pairing amplitude
for a wide range of initial states,
%% Periodic oscillations are observed in a wide range of couplings,
both for the initial states close to the normal state
($g_i\ll g$) as well as for the initial states near equilibrium ($g_i\approx g$).
At increasing $g_i$ there is a critical point at which
the asymptotic pairing amplitude $\Delta_\pm$ becomes zero.

%%%%%%%%%%%%%%%%%%%%%%%%%%%%%%%%%%%%%%%%%%%%%%%%%%%%%%%%%%%%%%%%%
\begin{figure}[t]
\includegraphics[width=3.5in]{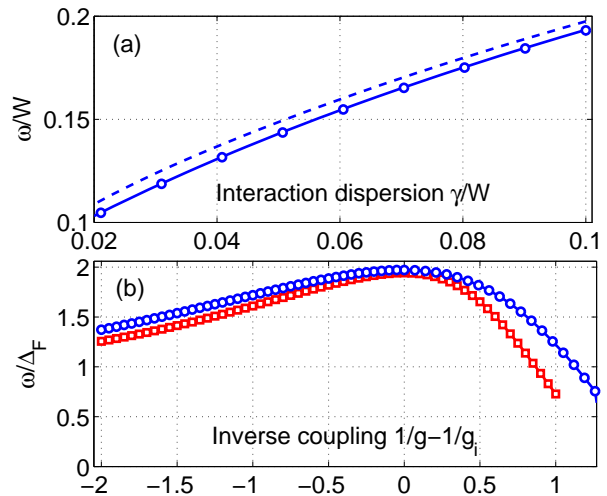}
%\vspace{-0.25cm}
\vspace{-0.8cm} \caption[]{(a): The Higgs mode frequency obtained from the
simulation with $g_i\approx g$ (circles) and from Eq.(\ref{eq:eigen_delta})
(solid line) as a function of the dispersion parameter $\gamma$. The
quasiparticle energy minimum (dashed line) lies above the collective mode
frequency (parameters of the simulation: $g=0.61$, $a_1=a_2=0.5$). (b):
Frequency of the Higgs mode as a function of the initial state for
non-separable and separable interactions with the same parameters as in
Fig.\ref{fig:A_regime}. The frequency changes away from $g_i=g$ as the
amplitude of oscillations increases (Fig.\ref{fig:A_regime}), indicating
unharmonicity of the Higgs mode. } \vspace{-0.3cm} \label{fig:coll}
\end{figure}
%%%%%%%%%%%%%%%%%%%%%%%%%%%%%%%%%%%%%%%%%%%%%%%%%%%%%%%%%%%%%%%%%%

To understand the origin of the oscillatory behavior for the dispersive
interaction, Eq.(\ref{eq:interaction}), we develop perturbation theory near the point
$g_i= g$. Linearizing the Bloch equations and taking a harmonic variation of the pairing amplitude,
$\delta \Delta^{x,y}_\vec p(t)\propto e^{-i\omega t}\delta \Delta^{x,y}_{\vec p\omega}$,
we find two collective modes for the $x$ and $y$ components of $\Delta_\vec p$
corresponding to the order parameter amplitude and phase variation (see Ref.\cite{Varma02}).
The amplitude (Higgs) mode with frequency $\omega$ obeys the integral equation
\be\label{eq:eigen_delta}
\delta \Delta^x_{\vec p\omega}=\frac12\sum_{\vec q}\frac{\lambda_{\vec p\vec q}\delta \Delta^x_{\vec q\omega}}{\sqrt{\epsilon_\vec q^2+\Delta^2_\vec q}}\,\frac{\epsilon^2_\vec q}{\epsilon_\vec q^2+\Delta^2_\vec q-\omega^2/4},
\ee
where
$\Delta_\vec p$ is the equilibrium gap obtained from Eq.(\ref{Eq:Gap_eq}).
The equation for $\delta\Delta^y_\vec p$ (the phase mode) is similar to
Eq.(\ref{eq:eigen_delta}) except for the denominator of the second fraction which is
$\epsilon_\vec q^2-\omega^2/4$. As expected from Goldstone theorem,
the equation for $\delta\Delta^y_\vec p$ is solved by $\omega=0$.

To find the frequency $\omega$ of the Higgs mode, we note that
for the interaction $\lambda_{\vec p\vec q}$ given by (\ref{eq:interaction}),
which is an operator of rank two, Eq.(\ref{eq:eigen_delta})
turns into an algebraic equation involving a $2\times2$ determinant.
Solving it we find that for $a_2>0$
the frequency $\omega$ lies within the BCS gap, as illustrated in Fig.\ref{fig:coll}a.
To gain more insight, let us consider a separable interaction, $a_1=0$, $a_2=1$, which
yields
\be\label{eq:delta_factorizable}
1=\frac{g}{2\nu_F}\sum_\vec q \frac{f^2_\vec q}{\sqrt{\epsilon_\vec q^2+\Delta_\vec q^2}}\,\frac{\epsilon^2_\vec q}{ \epsilon_\vec q^2+\Delta_\vec q^2-\omega^2/4},
\ee
where $\Delta_\vec q\propto f_\vec q$. Balancing the factors under the sum in order to obtain
unity on the left hand side, and noting that without the second factor
Eq.(\ref{eq:delta_factorizable}) would be identical to Eq.(\ref{Eq:Gap_eq}),
it is easy to see that $\omega<2\Delta_F$,
i.e. the Higgs mode is discrete.

Notably, as Fig.\ref{fig:coll}a illustrates,
the frequency obtained from Eq.(\ref{eq:eigen_delta})
coincides with the frequency of oscillations
in $\Delta_F(t)$ obtained by simulating BCS dynamics at $g\approx g_i$,
proving that the observed excitation is indeed the Higgs mode.
Furthermore, for $g$ away from $g_i$ the frequency extracted from  $\Delta_F(t)$
varies with $g$, decreasing below the value at $g\approx g_i$
and approaching zero at $g_i\ll g$ and $g_i\gg g$ (see Fig.\ref{fig:coll}b).
This indicates unharmonicity of the Higgs mode that sets on
at a large amplitude of oscillations.

To test these ideas further, we considered the regime when the Higgs mode
is strictly inside the quasiparticle continuum,
%%, $\omega>2\Delta$.
which can be realized in the model (\ref{eq:interaction}) with the second term
of a {\em repulsive} sign, $a_2<0$.
In this case Eq.(\ref{eq:eigen_delta}) has no real-valued solution
in the region $\omega\le 2\Delta$.
Simulating the BCS dynamics near $g_i\approx g$ we find that
$\Delta(t)$ exhibits exponentially decaying oscillations
of the form $e^{-\eta t}\cos(\omega't+\phi)$
corresponding to a complex-valued frequency $\omega$.
For $a_2=0$ the collective mode frequency $\omega=2\Delta_F$
lies at the edge of the quasiparticle continuum.
This property was linked to algebraic Landau damping
of this mode in Refs.\cite{Volkov74,Yuzbashyan05-2}.

\begin{figure}[t]
\includegraphics[width=3.5in]{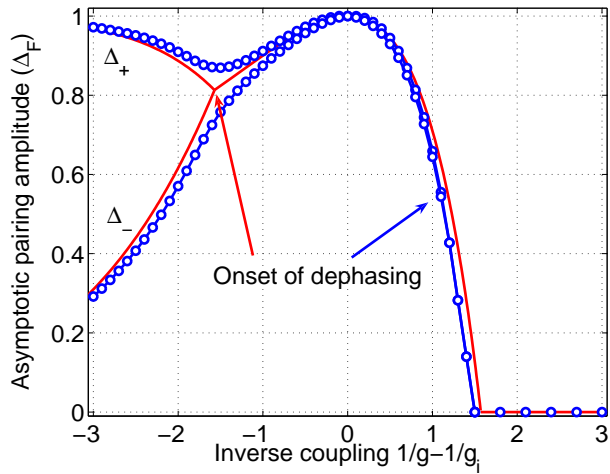}
%\vspace{-0.25cm}
\vspace{-0.8cm} \caption[]{Quenching of dephasing for weakly dispersing
interaction. Asymptotic values of the pairing amplitude $\Delta_\pm$ for a
dispersing (circles) and non-dispersing (red line) interaction for different
initial states. The onset of dephasing is marked by arrows. Parameters used:
$a_1=a_2=0.5$, $g=0.33$, $\Delta_F/W=0.02$, $\gamma/W=0.1$, and $a_1=1$,
$a_2=0$, $g=0.33$, $\Delta_F/W=0.05$, respectively. } \vspace{-0.3cm}
\label{fig:asymp_diag_wide}
\end{figure}

The discrete Higgs mode makes the BCS dynamics undamped for $g$ near $g_i$
even for weakly dispersing interaction $\lambda_{\vec p\vec q}$. It is interesting to
connect this behavior to the dephased BCS dynamics found in the case of constant
interaction. This is illustrated  (Fig.\ref{fig:asymp_diag_wide})
by the dynamics at weakly dispersing interaction
$\gamma\gg\Delta_F$, where we observe
%% We see
that the region of dephased dynamics shrinks, with
the onset of dephasing shifting towards small $g< g_i$.
While the oscillation amplitude $\frac12(\Delta_+-\Delta_-)$ is now finite,
it remains
small due to dephasing in the transient region
(see Fig.\ref{fig:A_regime}b).
This behavior is consistent with the Higgs mode approaching the quasiparticle
continuum boundary.

%%%%%%%%%%%%%%%%%%%%%%%%%%%%%%%%%%%%%%%%%%%%%%%%%%%%%%%%%%%%%%%%%
%%%%%%%%%%%%%%%%%%%%%%%%%%%%%%%%%%%%%%%%%%%%%%%%%%%%%%%%%%%%%%%%%%

In conclusion, we have shown that the energy dispersion of pairing interaction
leads to quenching of dephasing of the BCS dynamics, making the Higgs mode
of the pairing amplitude discrete. Parametric control
of interaction in the strong coupling regime near a Feshbach resonance of cold atoms
can be used to excite this mode.

This research was supported in part by the National Science Foundation under Grant No. PHY05-51164.


\vspace{-5mm}

\begin{thebibliography}{10}


%Observation of Resonance Condensation of Fermionic Atom Pairs
\bibitem{Regal04}%C. A. Regal, {\emph et al.}
C. A. Regal, M. Greiner, and D. S. Jin, Phys. Rev. Lett. {\bf 92}, 040403
(2004).

%Condensation of Pairs of Fermionic Atoms near a Feshbach Resonance
\bibitem{Zwierlein04}M. W. Zwierlein, C. A. Stan, C. H. Schunck, S. M. F. Raupach,
A. J. Kerman, and W. Ketterle, Phys. Rev. Lett. {\bf 92}, 120403 (2004).

% Theory of Feshbach resonance, a nice review
\bibitem{Timmermans99} E. Timmermans, P. Tommasini, M. Hussein,
and A. Kerman, Physics Reports {\bf 315}, 199 (1999).
%Feshbach resonances in atomic Bose-Einstein condensates


%Formation Dynamics of a Fermion Pair Condensate
\bibitem{Zwierlein05}M. W. Zwierlein, C. H. Schunck, C. A. Stan, S. M. F. Raupach, and W. Ketterle, Phys. Rev. Lett. {\bf 94}, 180401 (2005).


\bibitem{Dalfovo99}
F. Dalfovo, S. Giorgini, L. P. Pitaesvkii, and S. Stringari, Rev. Mod. Phys. {\bf 71}, 463 (1999).

\bibitem{Pethick_Smith}C. J. Pethick and H. Smith,
{\it Bose-Einstein condensation in dilute gases}, Cambridge University Press, 2002.


%Resonance Superfluidity in a Quantum Degenerate Fermi Gas
\bibitem{Holland01} M. Holland, S. J. J. M. F. Kokkelmans, M. L. Chiofalo, and R. Walser, Phys. Rev. Lett. {\bf 87}, 120406 (2001).

%Superfluidity and collective modes in a uniform gas of Fermi atoms with a Feshbach
%resonance
\bibitem{Ohashi03} Y. Ohashi and A. Griffin, Phys. Rev. A {\bf 67}, 063612 (2003).

%Universal Thermodynamics of Degenerate Quantum Gases in the Unitarity Limit
\bibitem{Ho04} Tin-Lun Ho, Phys. Rev. Lett. {\bf 92}, 090402 (2004).

%Effective Theory of Feshbach Resonances and Many-Body Properties of Fermi Gases
\bibitem{Bruun04} G. M. Bruun and C. J. Pethick, Phys. Rev. Lett. {\bf 92}, 140404 (2004).

%BCS-BEC Crossover: From High Temperature Superconductors to Ultracold Superfluids
\bibitem{Chen05} Q. Chen, J. Stajic, S. Tan, and K. Levin, Physics Reports {\bf 412}, 1 (2005).

%%Resonant Fermi Gases with a Large Effective Range
\bibitem{Schwenk05} A. Schwenk and C. J. Pethick, Phys. Rev. Lett. {\bf 95}, 160401 (2005).

%Measurement of the Interaction Energy near a Feshbach Resonance in a  6Li Fermi Gas
\bibitem{Bourdel03} T. Bourdel, J. Cubizolles, L. Khaykovich,  K. M.
Magalhães, S. J. Kokkelmans, G. V. Shlyapnikov, and C. Salomon,
Phys. Rev. Lett. {\bf 91}, 020402  (2003).

%Expansion of a Fermi Gas Interacting with a Bose-Einstein Condensate
\bibitem{Ferlaino04} F. Ferlaino, E. de Mirandes, G. Roati, G. Modugno, and M. Inguscio,
Phys. Rev. Lett. {\bf 92}, 140405 (2004).

%Damping of a Unitary Fermi Gas
\bibitem{Kinast05} J. Kinast, A. Turlapov, and J. E. Thomas, Phys.  Rev. Lett. {\bf 94},
170404 (2005).

%Potential Energy of a 40K Fermi Gas in the BCS-BEC Crossover
\bibitem{Stewart06} J. T. Stewart, J. P. Gaebler, C. A. Regal, and D.  S. Jin, Phys. Rev.
Lett. {\bf 97}, 220406 (2006).

\bibitem{Vaks61} V. G. Vaks, V. M. Galitskii, A. I. Larkin,
Zh. Eksp. Teor. Fiz. {\bf 41}, 1655 (1961) [Sov. Phys. JETP {\bf 14}, 1177 (1962)].

\bibitem{Bardasis61} A. Bardasis and J. R. Schrieffer, Phys. Rev. {\bf 121}, 1050 (1961).

%Higgs Boson in Superconductors
\bibitem{Varma02} C. M. Varma, J. Low Temp. Phys. {\bf 126}, 901 (2002).


\bibitem{Ivlev72} B. I. Ivlev, Pis'ma v Zh. Eksp. Teor. Fiz. {\bf 15}, 441 (1972)
[JETP Lett. {\bf 15}, 313 (1972)].

%Amplitude collective modes in superconductors and their coupling to charge-density waves
\bibitem{Littlewood82} P. B. Littlewood and C. M. Varma, Phys. Rev. B {\bf 26}, 4883 (1982).
\bibitem{Volkov74} A. F. Volkov and Sh. M. Kogan, Zh. Eksp. Teor. Fiz. {\bf 65}, 2038, (1973) [Sov. Phys. JETP {\bf 38}, 1018 (1974)].
%Colissionless relaxation of the energy gap in superconductors

\bibitem{Eliashberg60} G. M. Eliashberg, Zh. Eksp. Teor. Fiz. {\bf 38}, 966 (1960) [Sov. Phys. JETP {\bf 11}, 696 (1960)].
%% Interactions between electrons and lattice vibrations in a superconductor

\bibitem{Szymanska05} M. H. Szyma\'nska, B. D. Simons, and K. Burnett,
Phys. Rev. Lett. {\bf 94}, 170402 (2005).
%Dynamics of the BCS-BEC Crossover in a Degenerate Fermi Gas

%Collective Rabi Oscillations and Solitons in a Time-Dependent BCS Pairing Problem
\bibitem{Barankov04-1} R. A. Barankov, L. S. Levitov, and B. Z. Spivak,
Phys. Rev. Lett. {\bf 93}, 160401 (2004).


%Atom-Molecule Coexistence and Collective Dynamics Near a
%Feshbach Resonance of Cold Fermions
\bibitem{Andreev04} A. V. Andreev, V. Gurarie, and L. Radzihovsky, Phys. Rev. Lett. {\bf 93}, 130402 (2004); R. A. Barankov and L. S. Levitov, Phys. Rev. Lett. {\bf 93},
130403 (2004).


\bibitem{Warner05} G. L. Warner and A. J. Leggett, Phys. Rev. B {\bf 71}, 134514
(2005).
%Quench dynamics of a superfluid Fermi gas

\bibitem{Yuzbashyan05-2} E. A. Yuzbashyan, O. Tsyplyatyev, and B. L. Altshuler,
Phys. Rev. Lett. {\bf 96}, 097005 (2006); Erratum: Phys. Rev. Lett. {\bf 96}, 179905 (2006).

%Synchronization in the BCS Pairing Dynamics as a Critical Phenomenon
\bibitem{Barankov06-2} R. A. Barankov and L. S. Levitov, Phys. Rev. Lett. {\bf 96}, 230403 (2006).

% Dynamical Vanishing of the Order Parameter in a Fermionic Condensate
\bibitem{Yuzbashyan06} E. A. Yuzbashyan and M. Dzero, Phys. Rev. Lett. {\bf 96}, 230404 (2006).

\bibitem{AndersonMorel} P. Morel and P. W. Anderson, Phys. Rev. {\bf 125}, 1263 (1962).


\end{thebibliography}
\end{document}